\documentstyle[epsf]{l-aa}

\begin{document}

   \thesaurus{01         
              (11.04.1;  
               11.09.1;  
               12.07.1  
               )} %
   \title{The gravitational lens CFRS14.1311 = HST 14176+5226
         }

   \author{David Crampton \inst{1}\thanks{Visiting Astronomers, Canada-France-Hawaii Telescope, which
is operated by the National Research Council of Canada,
the Centre National de la Recherche Scientifique, and the University of Hawaii}, O. Le F\`evre \inst{2}$^{\star}$, F. Hammer \inst{2}$^{\star}$ \&
       S.J. Lilly \inst{3}$^{\star}$
          }

   \offprints{D. Crampton}

   \institute{Dominion Astrophysical Observatory, National Research Council of Canada, RR 5 Victoria, V8X 4M6 Canada.\\ Email: crampton@dao.nrc.ca 
    \and Observatoire de Meudon, Meudon, 92195, France
    \and Department of Astronomy, University of Toronto, Toronto, M5S 1A7
     Canada            }

   \date{Received 23 December 1995  / Accepted  30 January 1996}

   \maketitle
\markboth{Crampton et al.: CFRS14.1311 gravitational lens}{CFRS14.1311}

   \begin{abstract}

Ratnatunga et al. (1995) have recently proposed that an object, HST
14176+5226, in the ``Groth-Westphal" HST survey strip is an ``Einstein
cross'' gravitational lens.  By chance, this object has been previously
observed in the Canada-France Redshift Survey.  The candidate lens,
catalogued as CFRS14.1311, is an elliptical galaxy at z = 0.81.  In
addition, the spectrum shows a strong, spatially-extended, emission
feature at 5342\AA\ that almost certainly originates from two of the
four ``lensed'' images. We tentatively identify this emission line as
Ly $\alpha$ at z =3.4. A less prominent emission feature at
6822\AA\ may be C IV 1549.  Our data thus support the identification of
this system as a new quadruple-image lens.

      \keywords{galaxies: individual: CFRS14.1311 (= HST 14176+5226)
               -- cosmology: gravitational lensing
                               }
   \end{abstract}


\section{Introduction}

Ratnatunga et al. (1995; hereafter ROGI) have recently proposed that
two objects found in archival Hubble Space Telescope images are new
examples of ``Einstein cross'' gravitational lenses produced by distant
elliptical galaxies.  In each case four very faint blue compact images
are symmetrically placed around a brighter red galaxy.  One of these
objects, HST 14176+5226, is in the ``Groth-Westphal" GTO survey strip
(Groth et al. 1994). By chance, this object has been previously
observed during the Canada-France Redshift Survey (CFRS) as CFRS14.1311.  This
fact is not as improbable as it may seem, since the 1415+52 CFRS field
and the ``Groth-Westphal'' survey strip were both (and independently)
co-located with the ultradeep radio field of Fomalont et al.  (1991) to
allow identifications of these radio sources (see Hammer et al. 1995;
CFRS VII).

As well as pointing out the previously-published redshift for the
candidate lens (z = 0.81, Lilly et al. 1995b; CFRS III), we show in
this {\it Letter} that our spectrum of CFRS14.1311 also displays a
spatially extended emission line at 5342 \AA\ that almost certainly
originates from two of the four faint blue images. This line is likely
to be Ly $\alpha$ at z = 3.4. A less prominent emission feature at
6822\AA\ may be C IV at this redshift.  Our data thus strongly support
the identification of this system as a new quadruply-imaged
gravitational lens.  The secure redshift of the lens and the proposed
redshift of the lensed object will allow more detailed modelling of
this system.

\section{Observations}

Accurate astrometric positions for all objects in the CFRS 1415+52
field were determined to enable identification and investigation of the
optical counterparts of the $\mu$Jy radio sources catalogued by
Fomalont et al. (1991). The resulting optical positions agree with
those of their radio counterparts to within 0\farcs5 (CFRS VII). The
(2000) coordinates of CFRS14.1311 are 14$^h$ 17$^m$ 35\fs70 +52\degr
26\arcmin 46\farcs0, which differs from the position given by ROGI by
several arcseconds. However, it is clearly the same object and
Ratnatunga (private communication) agrees that the published 
ROGI position is in
error. No radio source is present at the location of CFRS14.1311 on the
VLA map (with S$_{5GHz}$ $>$ 8$\mu$Jy) published by Fomalont et al.
(1991).

Details of the CFRS photometric observations are given in Lilly et al.
(1995a; CFRS I), and the spectroscopic observations are described in Le
F\`evre et al. (1995; CFRS II).  The full 1415+52 field catalogue,
including CFRS14.1311, is presented in 
CFRS III.

\subsection{The lensing galaxy}

The candidate lens, CFRS14.1311, is an elliptical galaxy at 
z = 0.809.  Our published redshift of this galaxy in CFRS III is z = 0.807,  
but the displacement of the galaxy in the slit (see below) suggests that the
redshift was underestimated by 0.002. Koo et al. (1996) have
also observed this galaxy and find a very similar 
redshift, z = 0.811.

The spectrum of CFRS14.1311 is shown in Figure 1. The strong
``4000\AA\ break", the Ca H and K lines, and the G band unmistakeably
indicate the redshift for this galaxy.  Our isophotal $I$ magnitude, $I
= I_{AB} - 0.48 = 19.5$ is comparable to the total F814W magnitude,
19.7, reported by ROGI. The $(B-V)_{AB}$ colour, measured in a 3\arcsec\
aperture is 1.4 mag. With $H_0 = 50 $km s$^{-1}$Mpc$^{-1}$, the absolute
magnitude of this galaxy is M$_{AB}(B) = -22.9$ with $q_0 = 0.5$ and
M$_{AB}(B) = -23.4$ with $q_0 = 0.0$.


 \begin{figure}
     
      \epsfxsize=8.5cm \epsfbox{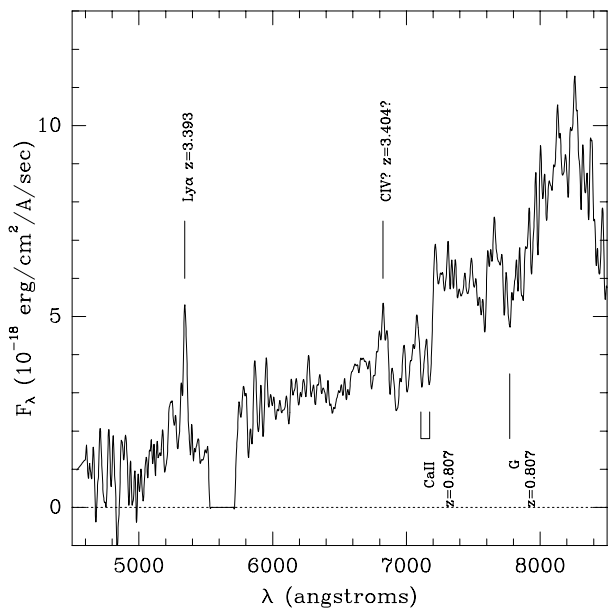}
      \caption[]{Spectrum of CFRS14.1311 showing the characteristic spectrum
        of an elliptical galaxy with absorption featues marked, plus two
        emission lines. The identification of the emission line at
      6822\AA\ with redshifted C IV is less certain than that of 5342\AA\ with
      Ly $\alpha$. The wavelength region 5500\AA -- 5700\AA\ 
        was obscured by an overlapping first order spectrum
        (see CFRS II).             }
        
   \end{figure}

\subsection{The lensed object}

As illustrated in Figure 1, the spectrum of CFRS14.1311 also shows 
emission features at 5342\AA\ and  6822\AA. The
line at 5342\AA\ is clearly seen on all seven spectra taken of this
object between 1993 Feb 23-25 and does not correspond to any known line
at the redshift of the elliptical galaxy ($\lambda_{rest} \sim 2956
\AA$).  The line at 6822\AA\ is less compelling as it is located in a
region of the elliptical galaxy spectrum ($\lambda_{rest} \sim 3775
\AA$) that has strong features produced by multiple stellar absorption
lines below the 4000 \AA\ break.

The identification of the 5342\AA\ emission feature with a background
source is strongly supported by the extent and orientation of the
emission line on our 2-dimensional spectrogram shown in 
Figure 2.  The left panel of Figure 2 shows the slit position for our
spectroscopic observations superimposed on a montage of our
ground-based image of CFRS14.1311 and the HST image (Figure 1 of ROGI
shows a nice HST image of the lens: the orientation of their figure is
such that North is diagonally up to the right, at $\sim59^{\circ}$ to
the vertical). {\it Post facto} examination of our spectroscopic data shows
unambiguously that our 1\farcs75 East-West slit was displaced 0\farcs55
northwards from the center of the elliptical galaxy.  This was
undoubtedly due to the fact that this particular slit was located near
the corner of the spectrograph slit mask where the astrometric
distortion corrections used in cutting the mask were less certain. This
displacement accounts for 0.002 of the redshift difference between our
previously published value and that of Koo et al. (1996).

   \begin{figure}[htbp]
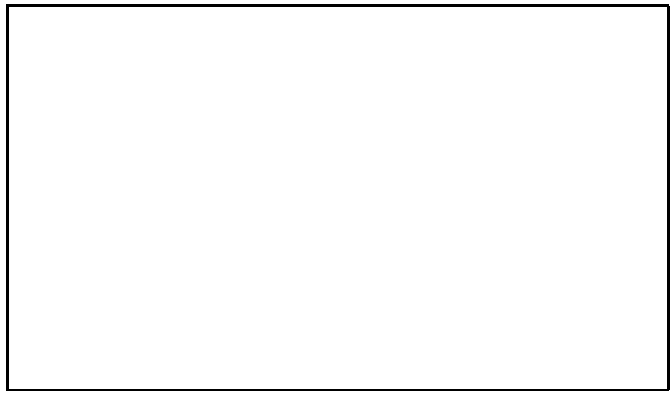

      \picplace{5.1cm}
      \caption{Left: superposition of HST (from ROGI)
       and ground-based images of
      CFRS14.1311 showing the location of the spectrograph slit which
     was 1\farcs75 wide. North is at the top, east to the left.  Right:
    portion of the two-dimensional spectrum showing the morphology of
    the Ly $\alpha$ emission line.  In this background-subtracted
    spectrum, wavelength increases upwards and east is to the left. The
   scales in each panel are 2\arcsec\ long.
              }
         \label{lensimage}
   \end{figure}

 As shown in Figure 2, both of the two northern images of the lens
would have been included in the slit. That the 5342 \AA\ emission
feature comes from these two images is indicated by the morphology of
the line on our two-dimensional spectrogram shown in the right panel.  The
line is noticeably more extended than the elliptical galaxy and the
extent, 2\arcsec, agrees well with the projected separation of the two
northern images.  Even more compelling, there is precise agreement
between the apparent velocity gradient across this emission line and
the spatial orientation of the two images if it is assumed that the
apparent velocity gradient arises purely from the relative displacement
of the images across the slit, i.e., the west side is shifted to higher
wavelengths, or North, by 0\farcs3.  Since both the spatial extent and
orientation of the 5342\AA\ emission line agree so precisely with that
expected from the image geometry, we believe the association of this
emission feature with the two northern images is secure.

The other emission feature is less convincing. It doesn't 
appear to be extended and its wavelength differs by 15\AA\ from the
expected position of C IV at the same redshift. However, it
is much weaker and is undoubtedly affected by the underlying
elliptical galaxy spectrum. We note that it is located almost precisely
at an apparent emission feature (actually due to surrounding absorption lines)
in elliptical galaxy spectra, but it is considerably stronger than usual.

The observed equivalent width of the 5342\AA\ emission feature is
55\AA.  Assuming a F606W magnitude difference of roughly 4.1
magnitudes (ROGI) between the elliptical galaxy and each of the four
quad images, this implies an observed equivalent width of roughly 1200\AA\
in each of the background objects.  This very high equivalent
width immediately suggests that the line is at high redshift, and, in
particular, that it is Ly $\alpha$ at z = 3.394, which would still
imply a rest-frame equivalent width of $\sim$270\AA.
The proposed Ly $\alpha$ emission feature is not resolved at our
resolution (projected slit = 35\AA) and hence is ``narrow'', raising
the possibility that it might be the result of a strong starburst
rather than an AGN.  

Some support for this redshift identification comes from the less
secure line at 6822\AA\ which is very close to the wavelength of C IV
1549 at this redshift, though it should be noted that this feature does
not appear to show the morphological signatures that made the
association of the 5342\AA\ line with the background objects so
compelling. If real, this would indicate that the background object is
a quasar, although the implied rest-frame equivalent width, 115\AA\
(but very uncertain) is uncomfortably high.  Less than 1\% of QSOs have
such strong lines (e.g. Hartwick \& Schade 1990, Crampton et al.
1990).  However, the absolute magnitude of the (lensed) object is M$_b$
= $-$23 for H$_0$ = 50 km s$^{-1}$ Mpc$^{-1}$, q$_0$ = 0, precisely at
the classical dividing line between quasars and Seyfert galaxies.
Large Ly $\alpha$ equivalent widths are more common among less luminous
quasars, an effect first noticed by Baldwin(1977). Since nearly all quasars are variable, it is also possible that the lensed images were brighter during our observations which were
taken 13 months earlier than the HST observations (1994 Mar 11).
Assuming an isothermal potential for the lens, simple models indicate
that the magnification of the faintest image in a quadrupole system
with the source nearly exactly aligned with the lens would be $\sim$2.
This would reduce the absolute magnitude of the background source to
about M$_b$ = $-$22 for H$_0$ = 50 km s$^{-1}$ Mpc$^{-1}$, q$_0$ = 0,
or  $\sim -$21 for q$_0$ = 0.5.  Thus the nature of the background
object remains more uncertain than the redshift, but the quality of our
data does not warrant further speculation on this point.

To summarize, we believe that the association of the emission feature
at 5342\AA\ with the northern pair of the quad pattern of images is
secure.  Identification of this line with Ly $\alpha$ at z = 3.394 is
then highly probable, based on the high equivalent width that is
implied.  If the C IV line is real, it would confirm the
redshift and would further imply that the background object is likely
to be a low luminosity quasar.  If the C IV is not real, however, then
the background object could be a less active AGN or a star-forming
galaxy.  In all cases, a redshift of z = 3.4 remains by far
the most plausible redshift, although lower redshifts can not be
categorically ruled out.

\section{Conclusion}

Our spectra thus indicate that the lensing elliptical galaxy CFRS 14.1311
has a redshift z = 0.809, and that the lensed object is a quasar or
a strong emission-line galaxy at z
= 3.4.  At these redshifts a simple singular isothermal lens model
using the parameters derived by ROGI implies a lens dispersion of
$\sigma = 230$ km s$^{-1}$  and a mass-to-light ratio M/L $=$
14h$^{-1}$ M$_{\sun}$/L$_{\sun}$. This indicates that CFRS14.1311 is
probably a straightforward case of a simple, isolated lens.

Although Refsdal and Surdej(1994) list eight four-image lenses in their
recent review, several of these have complex geometry of either the
lensed images and/or the lensing object(s). CFRS14.1311 appears to be
one of simpler cases involving an isolated normal elliptical galaxy
acting as a lens for a background AGN, and thus could, in principle,
provide cosmological constraints through further monitoring (e.g.,
Refsdal and Surdej 1994, Schneider 1995). We emphasize that
confirmation of our source redshift should be undertaken.

ROGI point out that their discovery rate of candidate lenses, two per
50 WFPC2 fields, is consistent with the surface density of bright
foreground elliptical galaxies and the surface density of $I < 26$
objects in the survey fields. Thus we might expect that the lensed
object should be broadly ``typical'' of the compact objects seen at $I
\sim 26$ in deep HST images.  If we assume that the lensed object
discussed here is indeed typical, then a large fraction of all of the
faint background objects (8000 per WFPC2 field) should have strong Ly
$\alpha$ emission.  However, neither the extensive searches that have
been undertaken for Ly $\alpha$ from primeval galaxies (e.g., Pritchet
1994), nor a simple extrapolation of the observed quasar number counts
(e.g., Schade et al.  1996) to $I = 26$, only $\sim$3000  deg$^{-2}$,
would support such a conclusion.

Finally,  this example of gravitationally
lensed objects also demonstrates that lensing effects must be taken
into account when deep HST images of field galaxies are being used to
derive conclusions about numbers of companions, merger rates, and
evolution in the high redshift universe.

\begin{acknowledgements}
The CFRS project has benefited from the support of the CFHT directors and
TACs, and from a NATO travel grant. We thank D. Duncan for making the mosaic
shown in Figure 2 and the Canadian Astronomy Data Centre for providing
archival material. We also acknowledge communication from
Marion Schmitz of the NED database project who independently noticed the
coincidence of HST 14176+5226 and CFRS 14.1311. 
     
\end{acknowledgements}

\end{document}